\documentclass[aps,tightenlines,twocolumn,floats,prd,nofootinbib,superscriptaddress,showpacs
]{revtex4-1}%%
\def\bea{\begin{eqnarray}}
\def\eea{\end{eqnarray}}
\def\bal{\begin{align}}
\def\eal{\end{align}}

\usepackage[dvips]{color}
\usepackage{graphics}
\usepackage{graphicx}
\usepackage{epsf} 
\usepackage{amsmath}
\usepackage{amssymb}
\usepackage{bbold}
\usepackage{bm}
\usepackage{slashed}
\usepackage{float}
\usepackage{psfrag}
\usepackage{footnote} 
\usepackage{xcolor}

\begin{document}

\hspace{5in}\parbox{1.5in}{ \leftline{JLAB-THY-13-1815, CFTP/13-024}
%                \leftline{WM-05-???}
%			             \leftline{nucl-th/05?????}
                \leftline{}\leftline{}\leftline{}\leftline{}
%\vspace{-3.6in}  % moves the preprint box down
}

\title
{\bf Pion electromagnetic form factor in the Covariant Spectator Theory 
}

\author{Elmar P. Biernat}
\affiliation{Centro de F\'isica Te\'orica de Part\'iculas (CFTP), 
Instituto Superior T\'ecnico (IST), Universidade de Lisboa, 
Av. Rovisco Pais, 1049-001 Lisboa, Portugal}
\author{Franz Gross }
 \affiliation{ Thomas Jefferson National Accelerator Facility (JLab), Newport News, VA 23606, USA}
\author{M. T. Pe\~na}
\affiliation{Centro de F\'isica Te\'orica de Part\'iculas (CFTP), 
Instituto Superior T\'ecnico (IST), Universidade de Lisboa, 
Av. Rovisco Pais, 1049-001 Lisboa, Portugal}
\author{Alfred Stadler}
  \affiliation{ Departamento de F\'isica da Universidade de \'Evora, 7000-671 \'Evora, Portugal \\and Centro de F\'isica Nuclear da Universidade de Lisboa (CFNUL), 1649-003 Lisboa, Portugal}
  \date{\today}

\begin{abstract}

The pion electromagnetic form factor at spacelike momentum transfer is calculated in relativistic impulse approximation using the Covariant Spectator Theory. The same dressed quark mass function and the equation for the pion bound-state vertex function as discussed in the companion paper are used for the calculation, together with a dressed quark current that satisfies the Ward-Takahashi identity.
The results obtained for the pion form factor are in agreement with experimental data, they exhibit the typical monopole behavior at high-momentum transfer, and they satisfy some remarkable scaling relations.
\pacs{11.15.Ex, 11.30.Rd, 12.38.Aw, 12.39.-x, 13.40.Gp, 14.40.Be}

\end{abstract}
\phantom{0}
\maketitle

\section{Introduction}
In the companion paper~\cite{mass_function_paper} (referred to as Ref.~I)  a model for the $q\bar q$ interaction was developed that uses the NJL mechanism to ensure that a pion bound state of zero mass exists whenever mass can be spontaneously generated through the self-interactions that dress a massless quark.  A novel feature of this model is its simplicity; in momentum space the kernel is the sum of a pure vector  $\delta$-function interaction and an interaction which provides confinement,  so that even though a feature of the Covariant Spectator Theory (CST)~\cite{Gro69,Gro74,Gro82} is that one of the quarks can be on-shell (where both the $q$ and $\bar q$ will sometimes be referred to collectively as ``quarks''), both quarks in the pair can never be on shell simultaneously.  The confining interaction can be a mixture of vector and scalar exchanges, but in the chiral limit (where the undressed mass of the quark, $m_0$, is zero) the scalar  part of the confining interaction decouples, allowing the 
chirally invariant 
vector 
interactions to preserve the features of chiral symmetry.   In Ref.~I the mass function was calculated by fitting two model parameters to lattice  data, and the bound-state $q\bar q$ equations were defined and their properties studied.  

It is the purpose of this paper to show that the simple model introduced and fixed in Ref.~I can be used to calculate the pion form factor without modifications.  This is the first demonstration showing how the model can be applied to a variety of interesting physics problems.  Even though the CST has been well studied, and used in previous calculations of nuclear form factors~\cite{Gross:2006fg,Pinto:2009dh}, this calculation introduces a number of new issues never before encountered. The discussion here will not only lead to some interesting new results, but also extend understanding of how to use the CST.  Discussion of the results, and comparison with some previous work, is saved for the last section.

%%%%%%%%%%%%%%%%%%%%%%%%%%%%%%%%%%%%%%

\section{Pion form factor in the Bethe-Salpeter theory}
\label{sec:piff}

\begin{figure}
\begin{center}
    \includegraphics[width=0.45\textwidth]{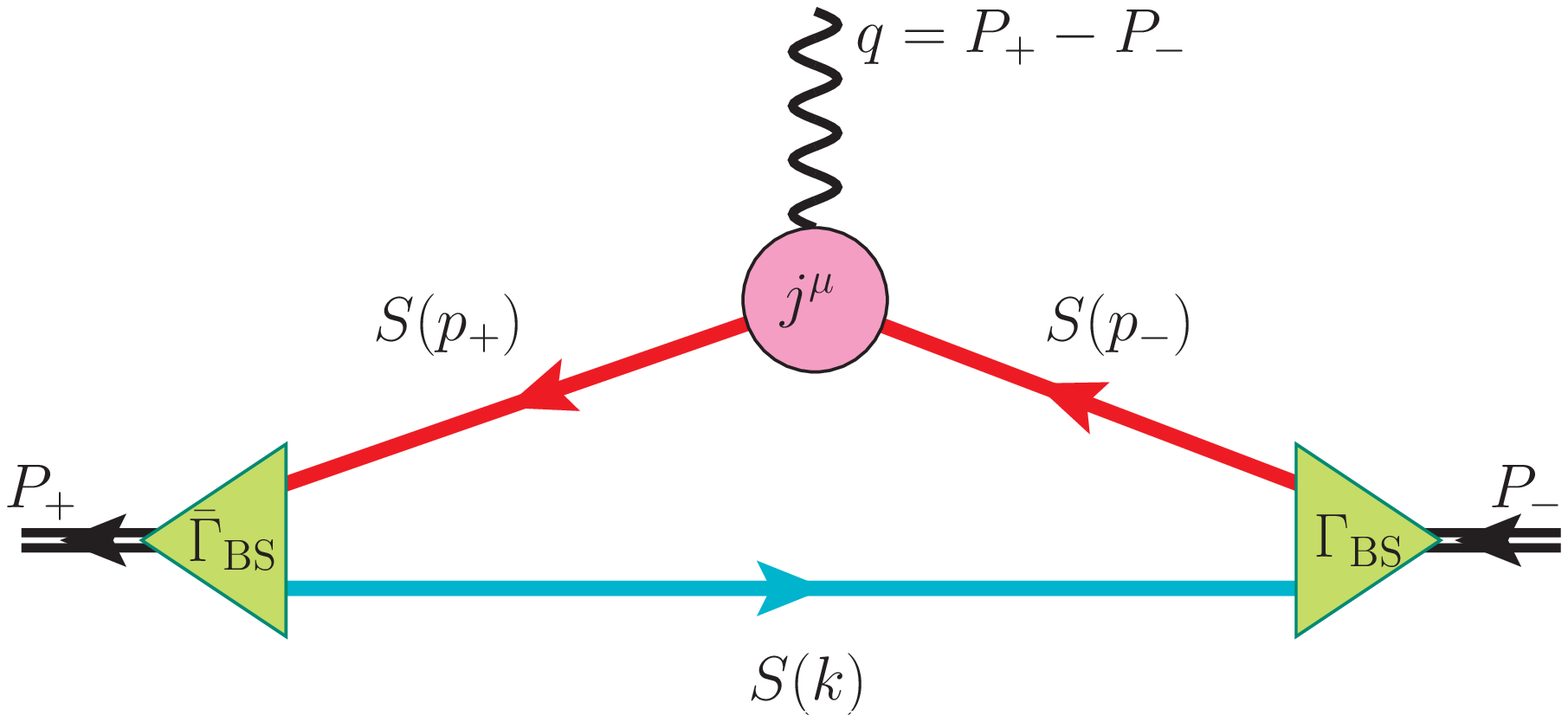}\vspace*{.5cm}
     \includegraphics[width=0.45\textwidth]{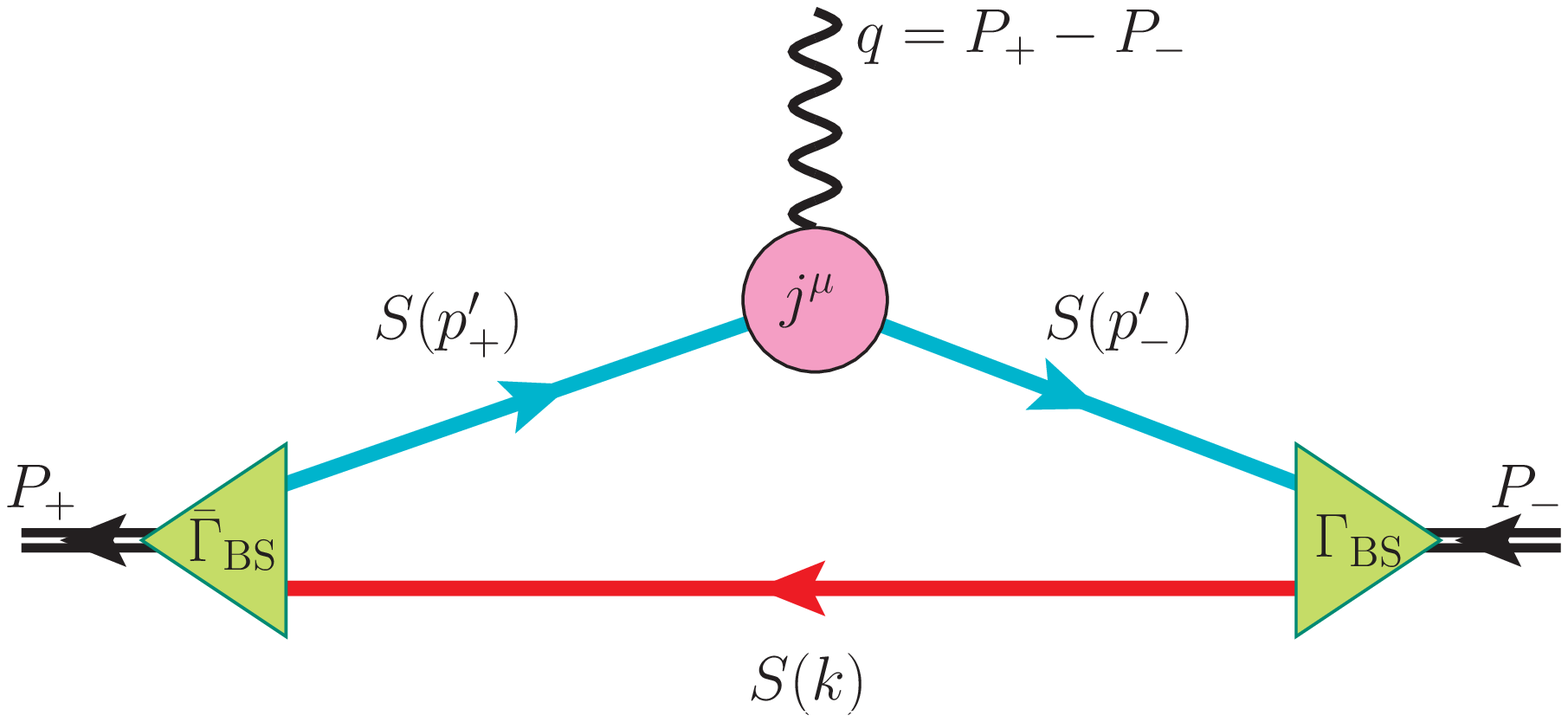}
% figure caption is below the figure
\caption{The two triangle diagrams for the electromagnetic pion form factor. Here $P_\pm$ are the outgoing and incoming (on-shell) pion four-momenta, $q$ is the four-momentum of the virtual photon, $S$ is the dressed quark propagator, $\Gamma_{\rm BS}$ is the Bethe-Salpeter pion vertex function and $j^\mu$ is the electromagnetic off-shell quark current. The top diagram describes the interaction of the virtual photon with the $u$ quark, with the $\bar d$ quark (represented by a $d$ quark traveling backward in time with momentum $k$) as a spectator; the bottom diagram represents the interaction of the virtual photon with the $\bar d$ quark (again represented by a $d$ quark traveling backward in time) with the $u$ quark as the spectator.}\label{fig:BStriangle}
\end{center} 
\end{figure}

The electromagnetic pion form factor in the spacelike region has been calculated in a great variety of different approaches, see, e.g. Refs.~\cite{Maris:2000,Chang:2013nia,Coester:2005cv,Brodsky:2007hb,deMelo:2005cy,Carbonell:2008tz,Biernat:2009my,Ebert:2005es,Masjuan:2008,Ananthanarayan:2012aa,Ananthanarayan:2012tt,Ananthanarayan:2013dpa,Troitsky:2013aa}. We begin by reviewing the discussion of the pion form  factor in the Bethe-Salpeter (BS) formalism.  We will consider a positively charged $\pi^+$ consisting of a $u$ and a $\bar d$ quark; the form factor for the $\pi^-$ can be obtained by charge conjugation. In impulse approximation, the electromagnetic form factor of the $\pi^+$  is extracted from the sum of two triangle diagrams, in which the photon couples either to the $u$ or the $\bar d$ quark, as depicted in Fig.~\ref{fig:BStriangle}. 

The top diagram, with the $\bar d$ quark as spectator, is weighted by the $u$-quark's electric charge $\frac23\mathrm e$, while the bottom diagram, with the $u$ quark as spectator, is weighted by the electric charge $-\frac13\mathrm e$ of the $d$ quark traveling backward in time.  The sum of the two diagrams is
\begin{widetext}
\begin{eqnarray}\label{eq:BSpicurrent}
 J^\mu(P_+,P_-)=\mathrm e F_\pi (Q^2) (P_++P_-)^\mu
% \nonumber\\
&=&\frac23\mathrm e\int \frac{\mathrm d^4k}{(2\pi)^4}\,
\mathrm {tr}\Big[\overline{\Gamma}_{\rm BS}(k,p_+) S(p_+) 
j^\mu (p_+,p_-) S(p_-) \Gamma_{\rm BS}(p_-, k)S(k)\Big]
\nonumber\\&&
-\frac13\mathrm e\int \frac{\mathrm d^4k}{(2\pi)^4}\,
\mathrm {tr}\Big[\Gamma_{\rm BS}(k,p_-') S(p_-')
%\nonumber\\&& \times 
 j^\mu (p_-',p_+')S(p_+') \overline{\Gamma}_{\rm BS}(p_+',k)S(k)\Big]\,, 
\end{eqnarray}
\end{widetext}
where $p_\pm=k+P_\pm$, $p'_\pm=k-P_\pm$,  $j^\mu (p_+,p_-)$ is the dressed current for off-shell quarks  (defined below), and $S(p)$ is the dressed propagator of a quark with momentum $p$
\begin{eqnarray}
S(p)=Z(p^2)\frac{M(p^2)+\slashed p}{M^2(p^2)-p^2-\mathrm i \epsilon} \, ,
\end{eqnarray} 
with $M(p^2)$ being the quark mass function and $Z(p^2)$ the wave function renormalization, as discussed in Ref.~I and reviewed below.   For the model considered here, $Z(p^2)=1$. 
The quark mass function $M(p^2)$ is obtained from the solution of the CST Dyson equation for the self-energy, and the constituent (dressed) mass $m$ of the quark is then determined from the condition $M(m^2)=m$.   We assume equal masses for the $u$- and $d$-quarks, so the $u$ and $d$ propagators are identical.
Finally, following the notation of Ref.~I, the  vertex function $\Gamma_{\rm BS}(p_1,p_2)$ describes a $\pi^+$ coupling to an {\it outgoing\/} $u$ quark of momentum $p_1$ and an {\it incoming\/} $d$ quark of momentum $p_2$ (the same as an {\it outgoing\/} $\bar d$ quark of momentum $-p_2$), while $\overline{\Gamma}_{\rm BS}(p_1,p_2)$ describes a  $\pi^+$ coupling to an {\it incoming\/} $u$ quark of momentum $p_1$ and an {\it outgoing\/} $d$ quark of momentum $p_2$ (the same as an {\it incoming\/} $\bar d$ quark of momentum $-p_2$).

Before turning to the CST formalism, we show how the second contribution to the form factor can be transformed into the first, and the two added together. To do this we need the following transformations of the vertex function and the current under charge conjugation 
\bea
{\cal C}\Gamma^{T}_{\rm BS}(p_1,p_2){\cal C}^{-1}&=&\Gamma_{\rm BS}(-p_2,-p_1)  
\nonumber\\
 {\cal C}j^{\mu {T}}(p',p){\cal C}^{-1}&=&-j^\mu(-p,-p')\, . 
\label{eq:Ctrans}
\eea
While these relations can be derived from general principles, they also follow from the typical matrix structure of $\Gamma_{\rm BS}$ [such as $\Gamma_{\rm BS}(p_1,p_2)\sim (m-\slashed{p}_1)\gamma^5(m-\slashed{p}_2)$] or $j^\mu$ [such as $j^\mu(p',p)\sim (m-\slashed{p}')\gamma^\mu(m-\slashed{p})$].  Taking the transpose of the trace, inserting ${\cal C}{\cal C}^{-1}=1$ between the operators, and using the properties (\ref{eq:Ctrans}) converts the trace from second term of Eq.~(\ref{eq:BSpicurrent}) into
\bea
\mathrm {tr}\Big[\cdots\Big]&\to& -\mathrm {tr}\Big[\overline{\Gamma}_{\rm BS}(-k,-p_+')S(-p_+')
%\nonumber\\&& \times 
 j^\mu (-p_+',-p_-')
 \nonumber\\
 &&\qquad\times S(-p_-')\Gamma_{\rm BS}(-p_-',-k) S(-k)\Big]\, .
 \label{eq:tracebp}
\eea
Now changing $k\to -k$ under the integral, and noting that this converts $p_\pm'\to -p_\pm$, shows that the trace (\ref{eq:tracebp}) is identical to the trace in the first term of (\ref{eq:BSpicurrent}), except for the minus sign which converts the factor $-\frac13\to \frac13$.  Hence the two terms are identical (except for the charges) and their sum equals the first term with the factor of $\frac23 \mathrm e$ replaced by $\mathrm e$. Note that the ability to change $k\to-k$ under the integral was essential to the argument.  

This discussion can be easily extended to show that the $\pi^0$ form factor is identically zero, and that except for a sign, the $\pi^-$ form factor is identical to the $\pi^+$ form factor.
 
 %%%%%%%%%%% 
%%%%%%%%%%%

\section{Pion form factor in the CST} \label{sec:FFandCST}

In the CST, the integral over the relative momentum of the two propagating particles is constrained by the requirement that one of the two particles must always be on shell, with contributions from terms when both particles are off-shell moved to higher order in the series of terms that define the relativistic two-body kernel.  The motivation for this rearrangement of terms is that, in many examples, it can be shown that the off-shell terms (from box diagrams, for example) tend to cancel other higher order terms in the kernel (crossed box diagrams, for example) so that keeping one particle on-shell not only simplifies the equations, but also improves the convergence of the approximation to the underlying field theory.  

The pion form factor in CST will therefore also involve triangle diagrams similar to the BS diagrams shown in Fig.~\ref{fig:BStriangle}, but with the internal particles constrained to their mass shell in the same way that they are constrained in the two-body bound-state equation.  As shown in Ref.~I, a careful treatment of the pion bound state in the chiral limit requires a four-channel equation, with contributions from the positive and negative energy poles of both particles included.  The full treatment of the triangle diagram for the $\pi^+$ form factor using the four-channel equation would therefore involve contributions from the positive- and negative-energy poles of both the $u$ and $\bar d$ quark.

\begin{figure}[t]
\begin{center}
    \includegraphics[width=0.45\textwidth]{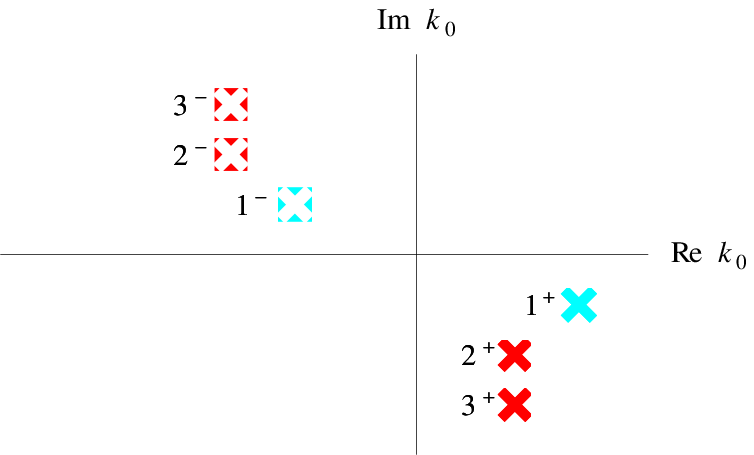} \\  
    \vspace{0.5cm}
    \includegraphics[width=0.45\textwidth]{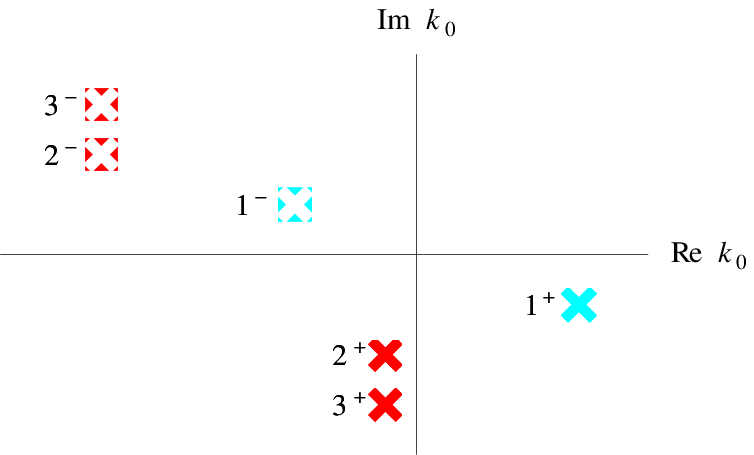}   
% figure caption is below the figure
\caption{The locations of the six propagator poles in the complex $k_0$-plane of the diagram where the $d$-quark is spectator, shown here for both $Q$ and $|\bf  k|$ small, with  $m=0.308$ GeV.  The top panel shows the case when $\mu=0.14$ GeV; the bottom shows the case when $\mu=0.42$ GeV. Note that large and different imaginary parts $\epsilon$ have been chosen for each pole  in order to spread them out in the complex plane for better illustration, but in all cases $\epsilon\to0$ is implied.}\label{fig:polesTriangle}
\end{center} 
\end{figure}

In this first calculation of the pion form factor using CST, we chose to make some approximations that still preserve the important physics.  To understand these approximations, study the location of the particle poles in the BS diagrams of Eq.~(\ref{eq:BSpicurrent}) and Fig.~\ref{fig:BStriangle}.  First concentrate on the diagram where the photon couples to the $u$-quark and the $\bar d$-quark is spectator (top panel in Fig.~\ref{fig:BStriangle}). This diagram has six propagator poles in the complex $k_0$-plane, three of them in the lower- and three in the upper-half plane \cite{Gro83}. In the Breit frame,
where
\bea
P_\pm&=&\{P_0,{\bf 0},\pm\frac12 Q\} 
\nonumber\\
q&=&\{0,{\bf 0},Q\} \label{eq:BreitFrame}
\eea 
with $P_0=\sqrt{\mu^2+\frac14Q^2}$,  $\mu$ the pion mass and $Q$ the photon momentum transfer, the poles of the spectator $d$-quark are located at
$k_0=\pm E_k\mp \mathrm i \epsilon$, denoted $1^{\pm}$, where $E_k = (m^2+{\bf k}^2)^{1/2}$, and the poles of the struck $u$-quark with momenta $p_-$ and $p_+$ are at 
\bea
k_0&=& -P_0\pm \sqrt{m^2+{\bf k}_\perp^2+\left(k_z-\frac Q2\right)^2}\mp \mathrm i \epsilon \, ,
\nonumber\\
 k_0&=&-P_0\pm \sqrt{m^2+{\bf k}_\perp^2+\left(k_z+\frac Q2\right)^2}\mp \mathrm i \epsilon \, ,
 \label{eq:k0pp}
 \eea
denoted $2^{\pm}$ and $3^{\pm}$, respectively. Since the square roots in the last two expressions are positive, recalling that $p_\pm=k+P_\pm$ means that $2^+$ and $3^+$ are the positive-energy poles and $2^-$ and $3^-$ are the negative-energy poles of the struck $u$-quark. The locations of these six poles in the complex $k_0$-plane depend on $m$, $\mu$, $Q$, $k_z$ and ${\bf k}_\perp=(k_x,k_y)$, and are shown in Fig.~\ref{fig:polesTriangle} for two different pion masses.  

\begin{figure}[t]
\begin{center}
    \includegraphics[width=0.45\textwidth]{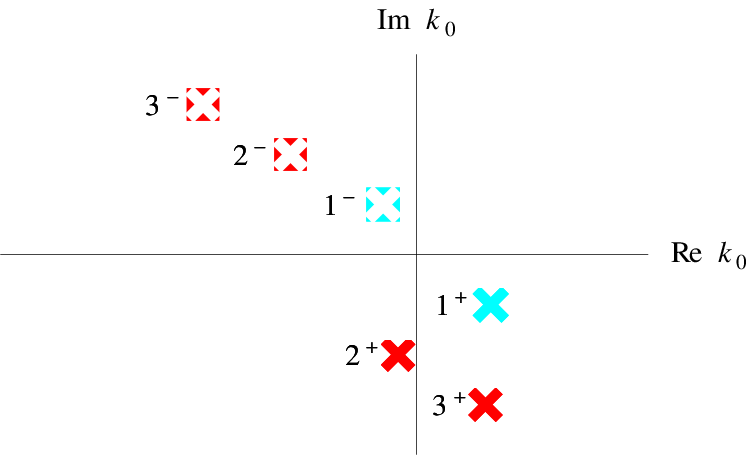}  
    \\  
    \vspace{0.5cm}
    \includegraphics[width=0.45\textwidth]{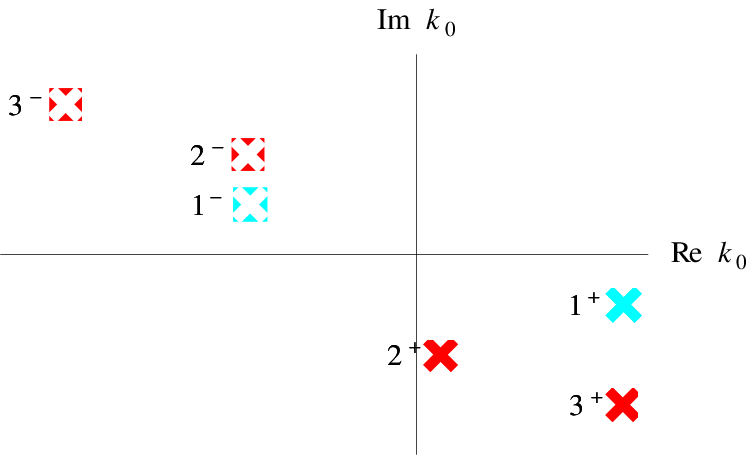}   
% figure caption is below the figure
\caption{The locations of the same six propagator poles shown in Fig.~\ref{fig:polesTriangle}, with  $m=0.308$ GeV,  $\mu=0.14$ GeV, and  ${k}_\perp=0$, but with $Q$ large.   The top panel shows the case when $k_z\lesssim Q/2$ and the poles $1^-$ and $2^+$ pinch;  
the lower panel shows the case when $k_z\gtrsim Q/2$ and  the poles $1^-$ and $2^-$ get close to each other.   }\label{fig:polesTriangle3}
\end{center} 
\end{figure}

Before proceeding further, recall that the masses in the denominators of the propagators are not fixed, but are functions of the four-momenta, so that, for example, the denominator of the spectator propagator is $M^2(k^2)-k^2$ not $m^2-k^2$.  {\it At the pole\/}, however, the mass condition $M(m^2)=m$ holds, so that the location and movement of the poles can be computed just as if the masses were fixed.

The full four-channel CST equation requires averaging  the contributions from all of  the propagator poles in the upper  and lower half planes.  Study of the bottom panel of Fig.~\ref{fig:polesTriangle} shows that, when $\mu$ is comparable to the dressed quark mass $m$,  the largest contribution will come from the $1^-$ pole.  For small $Q$ (and small $|\bf k|$) it is close to the poles at $2^+$ and $3^+$ in the lower half plane, and at the same time far away from the other poles.  This is the on-shell contribution of the spectator, with the physical energy of the outgoing $\bar d$ antiquark in its {\it positive\/}-energy state (because the incoming $d$ quark is in its {\it negative\/} energy state). This approximation, used previously in the study of deuteron form factors, is known as the relativistic impulse approximation (RIA)~\cite{Gross:1965zz,Arn77,Arn80,VO95}.

The top panel of Fig.~\ref{fig:polesTriangle} shows that for small $\mu$ (and also small $Q$) all of the poles in the upper-half plane are close to each other (and will coalesce into a triple pole when both $Q=0$ and $\mu=0$).  The requirement that  the limit $\mu\to0$  be described correctly is precisely what led to the need for a four-channel CST equation in the first place, and these additional channels, included in  contributions from the $2^-$ and $3^-$ poles, are also needed for a correct description of the form factor  in the limit when {\it both\/} $\mu$ and  $Q$ are small.  

In this first calculation, we will use the RIA, and hence we cannot expect to be able to correctly describe the form factor in the limit when both $\mu$ and $Q$ are small, where the neglected contributions from the $2^-$ and $3^-$ poles cannot be ignored (in fact, the RIA becomes singular when both $Q$ and $\mu$ tend to zero).

The case when $Q$ is large poses an interesting issue.  In this case the position of the poles is quite insensitive to the value of $\mu$, and therefore the RIA describes the form factor equally well for both large and small $\mu$.  
However, as $k_z\to Q/2$, the integrand becomes large, with the precise role of the poles depending on whether or not $k_z$ is less than or greater than $Q/2$ [recall Eq.~(\ref{eq:k0pp})].  If $k_z
\lesssim Q/2$ the poles $1^-$ and $2^+$ pinch, as shown in the top panel of Fig.~\ref{fig:polesTriangle3}, while if  $k_z\gtrsim Q/2$, the poles $1^-$ and $2^-$  are close together, as illustrated in the bottom panel of Fig.~\ref{fig:polesTriangle3}.  In both cases it looks like the integral could be singular, but it remains finite (and small).  %; for some background, see the discussion in Re.~\cite{Gro83}.  
Briefly, to see what is happening, it is necessary to examine the behavior of the propagator with momentum $p_-$ when the residue of the pole $1^-$ is evaluated, i.e., when the spectator is on its negative-energy mass shell, with $k_0=-E_k+i\epsilon$.
The relevant integral is 
\bea
I&\sim& \int dk_z\,f(k_z) [m^2 - p_-^2]^{-1}
\nonumber\\
&=&\int dk_z\, f(k_z) [-\mu^2+2P_0E_k-Qk_z]^{-1} \, ,
\eea
where we have approximated $M^2(p_-^2)\simeq m^2$ because we are interested in the kinematics where $p_-^2$ is close to $m^2$, and $f(k_z)$ is the remainder of the integrand which provides the needed convergence when $k_z\to \infty$.
As $Q$ becomes very large, this integral peaks at very large $k_z$, but is still finite.  To estimate it expand the factors
\bea
\lim_{Q\to\infty}I &\to&\int dk_z\frac{f(k_z)}{Qk_z} \Bigg[\Big(1+\frac{2\mu^2}{Q^2}\Big)\Big(1+\frac{E_\perp^2}{2k_z^2}\Big)-1\Bigg]^{-1}
\nonumber\\
&\simeq&\int dk_z\, f(k_z)\Bigg[\frac{2\mu^2k_z}{Q}+\frac{E_\perp^2Q}{2k_z} -\mu^2\Bigg]^{-1} \, ,
\label{eq:limI}
\eea
where $E_\perp=\sqrt{m^2+k_\perp^2}$.  
This shows that the integrand peaks at $k_z=E_\perp Q/(2\mu)$, and it is finite there provided that $\mu < 2m$. Because the peak is located at large values of $k_z$ where the remainder of the integrand, $f(k_z)$, is already very small, the integral is small as well.  
The rapid peaking at high $Q$ plays a crucial role in giving the correct asymptotic behavior of the form factor, as will be discussed in Sec.~\ref{sec:highQ2limit}  below.  Examination of the other propagator in $p_+^2$ shows a similar behavior, but at negative $k_z$.

To summarize, for small $Q^2$ the RIA, by retaining only the spectator pole contribution $1^-$, is a good approximation to the CST triangle diagram only for sufficiently large pion masses. For large $Q^2$, on the other hand, the locations of the poles are insensitive to $\mu$ and therefore the RIA is good not only for large but also for small values of $\mu$ (the  physical pion mass of $\mu=0.14$ GeV, for example) and even for vanishing pion mass in the chiral limit. 
This concludes our discussion of the RIA contribution from the spectator $d$ quark.  

 \begin{figure}[t]
\begin{center}
\includegraphics[width=0.45\textwidth]{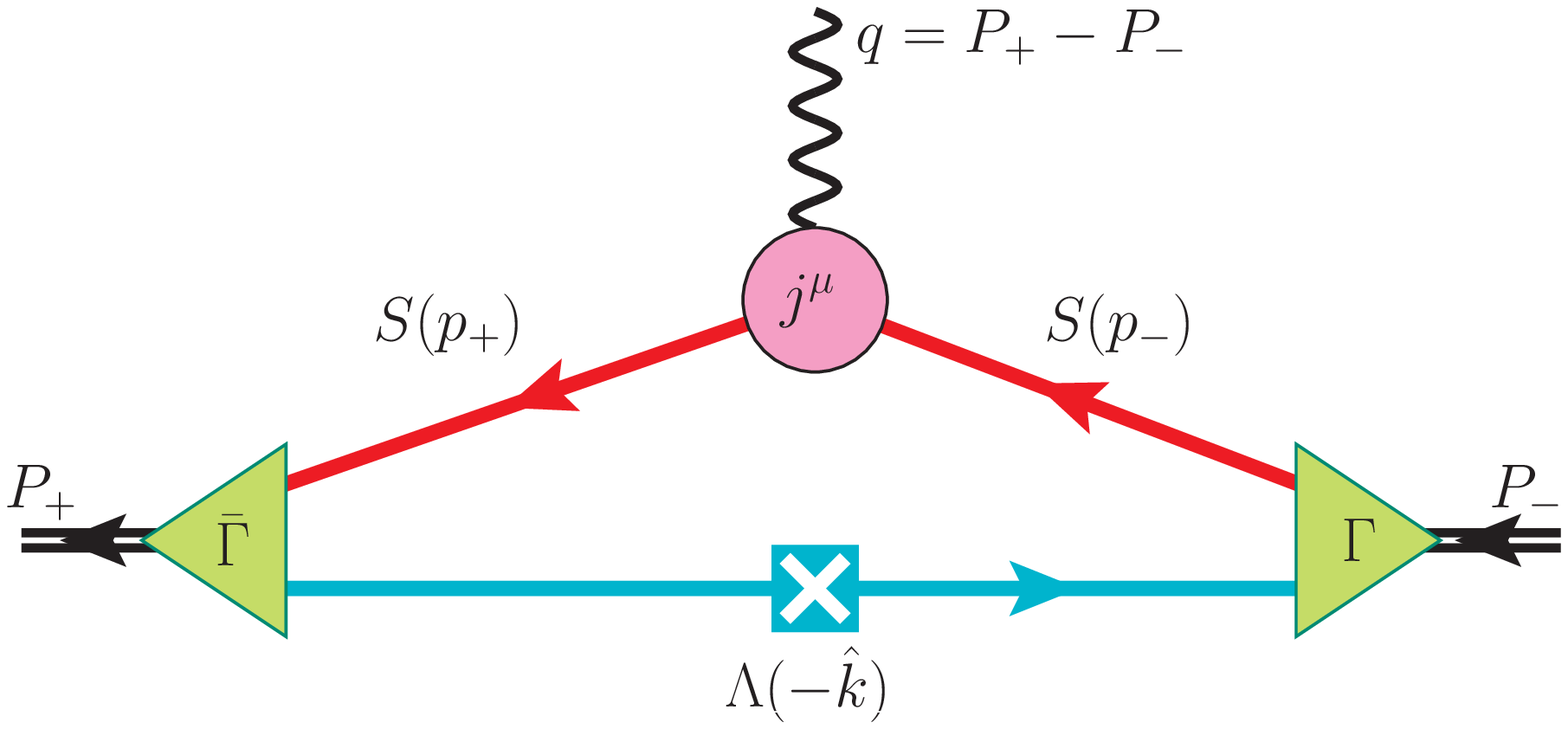}\vspace*{.5cm}
\includegraphics[width=0.45\textwidth]{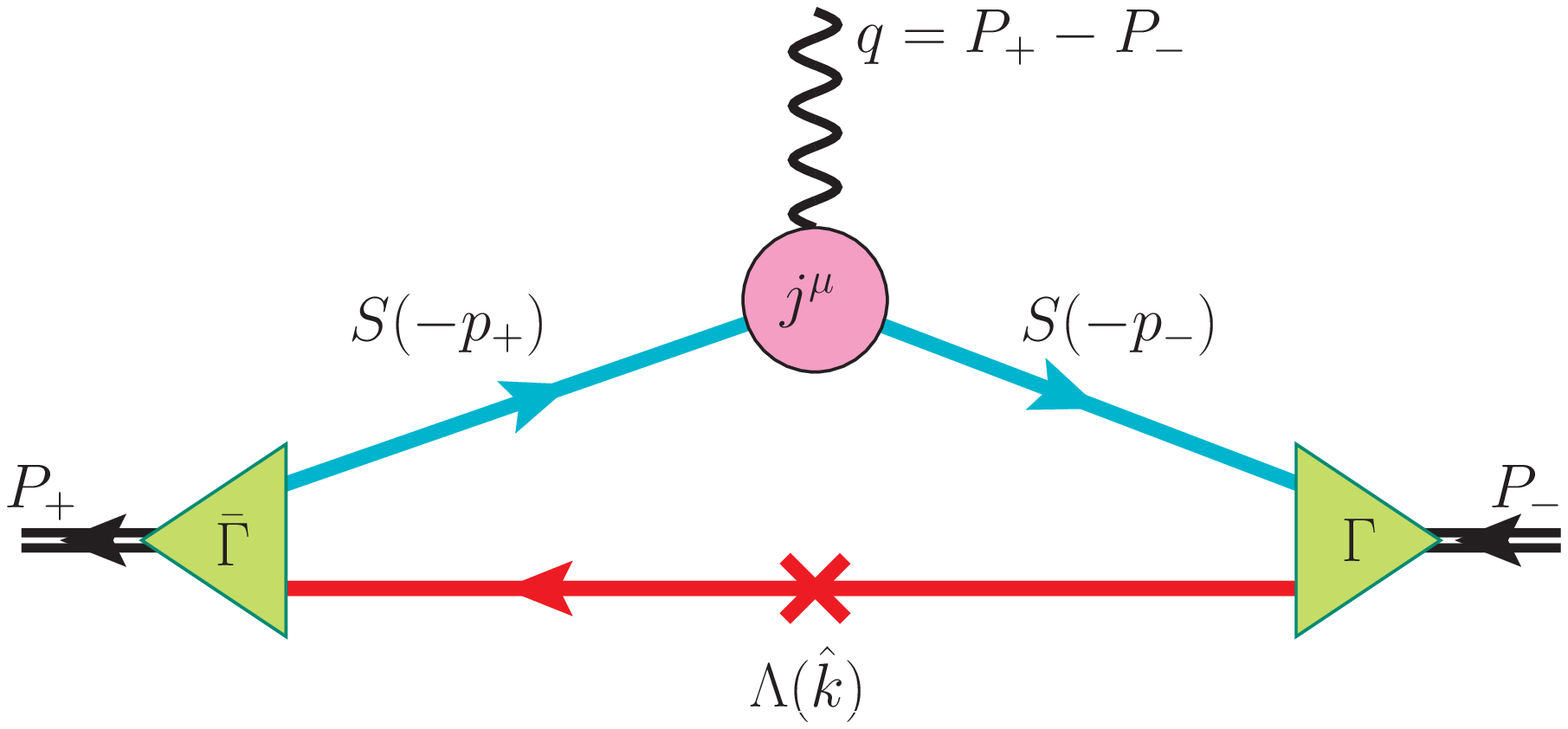} 
 % figure caption is below the figure
\caption{The two contributions to the $\pi^+$ form factor in the RIA.}\label{fig:polesTriangleRIA}
\end{center} 
\end{figure}

Now we turn to the RIA contribution from the diagram where the $u$-quark is the spectator. The locations of the poles can be analyzed in the same way as in the first case; this will not be discussed in detail here. In essence, it is now the positive-energy pole of the spectator $u$-quark (in the lower-half $k_0$-plane) that plays the same role as the spectator pole from the $d$-quark discussed above. This contribution is  represented diagrammatically in the lower panel of Fig.~\ref{fig:polesTriangleRIA}.

The diagram shown in the lower panel of Fig.~\ref{fig:polesTriangleRIA} can be transformed into the  expression for the upper panel, except for a different charge factor.
As in the BS case, invariance under the transformation $\hat k \rightarrow -\hat k$, where $\hat k=\{E_k,{\bf k}\}$ is the on-shell spectator quark momentum, is needed for this transformation, and it is possible because the first contribution is obtained from  the \emph{negative-energy} pole contribution of the spectator $d$-quark of the upper-half plane and the second from the \emph{positive-energy} pole contribution of the spectator $u$-quark of the lower-half plane. The first fixes $k_0=-E_k$ and the second $k_0=E_k$, and together with a change of the integration variable ${\bf k}\to -{\bf k}$ give the symmetry needed to relate the contributions from the $u$ and the $d$ spectators.   

Adding the two contributions yields the $\pi^+$ form factor in RIA,
\begin{eqnarray}\label{eq:picurrentA}
 J_{\rm RIA}^\mu(P_+,P_-)
&=&\mathrm e\,\int_k\,\mathrm {tr}\Big[
\bar\Gamma (-\hat k,p_+) S(p_+)  j^\mu (p_+,p_-)\nonumber\\&&
\times S(p_-) \Gamma(p_-, -\hat k) \Lambda(-\hat k)\Big] \, .
\end{eqnarray}   
Here, $p_\pm=P_\pm-\hat k$ are the off-shell quark momenta, $\Gamma(p,-\hat k)$ is the CST pion vertex functions, $\Lambda(\hat k)$ is the on-shell projector
\bea
\Lambda(\hat k)=\frac{m+\hat k}{2m} \, ,
\eea
and
the shorthand
\bea
\int_k\equiv\int\frac{ d^3k}{(2\pi)^3}\frac{m}{E_k}\, . \label{eq:kint}
\eea
is used for the momentum integration.

\section{Ingredients}

In this section the terms needed for the evaluation of the trace in Eq.~(\ref{eq:picurrentA}) are assembled.  In the next section the trace is evaluated  and its behavior 
as $Q\to\infty$ is examined.  Numerical results for the form factor are presented in Sec.~\ref{sec:Resuts}.

\subsection{Pion vertex function}
  
The pion vertex function $\Gamma(p,k)$ is required for the calculation of the trace in Eq.~(\ref{eq:picurrentA}).  The complete CST pion vertex function is obtained by solving the full four-channel pion bound-state equation of Ref.~I. 
This more ambitious task  will  be the subject of future work. Here we use an approximate pion vertex function that is an \emph{off-shell extension} near the chiral limit. Since the linear confining interaction does not contribute to the pseudoscalar bound-state equation in the chiral limit, as explained in Ref.~I, this estimate will be made using the vector  interaction only,  with the general form
\bea
{\cal V}_V=\frac14 V_C\,\gamma^\mu\otimes\gamma_\mu \, ,
\eea
where the specific form of  the scalar function $V_C$ will be given below.  
The CST equation for the bound state  (but with both external particles off shell) was already derived in Ref.~I; here we present an alternative derivation starting from the BS bound state  equation written in the rest frame in the chiral limit (where $P=0$)
\bea
\Gamma(p,p)=G(p^2)\gamma^5=i\int_{\rm poles}\frac{d^4k}{(2\pi)^4} {V}_C G(k^2) \frac{N}{D}\qquad
\label{eq:GinBSform}
\eea
where $G(p^2)$ is a scalar function, and the notation on the integral reminds us that the $k_0$ part of the integral is to be evaluated keeping {\it only\/} the $k_0$ poles from the quark propagators (and forsaking all others), and also anticipates the result in the chiral limit, where only the $\gamma^5$ structure will contribute to $\Gamma$.  The numerator in the chiral limit therefore becomes
\bea
N&=&\frac14 \gamma^\mu (M_\chi(k^2)+\slashed{k})\gamma^5(M_\chi(k^2)+\slashed{k})\gamma_\mu
\nonumber\\
&=&-(M_\chi^2(k^2)-k^2)\gamma^5
\eea
where $M_\chi(k^2)$ is the running mass function in the chiral limit (i.e. with $m_0=0$).  The denominator is
\bea
D=(M^2_\chi(k^2)-k^2-i\epsilon)^2\, .
\eea
To obtain the CST equation we are instructed to take the poles of the propagators {\it only\/}, which, after the cancellation of the factor $M^2_\chi(k^2)-k^2$ are single poles at $k_0=\pm E_k =\pm\sqrt{m_\chi^2+{\bf k}^2}$ with $m_\chi$ the root of the mass equation in the chiral limit,  $m_\chi=M_\chi(m_\chi^2)$.  Now that one of the initial quarks is on-shell, it is possible to specify the scalar function $V_C$.  

As discussed in Ref.~I,  in the general case when the total four-momentum $P$ may not be zero, 
\bea
{V}_C(p_1,p_2;k_1,\hat k_2)&=& 2C\, \frac{E_k}{m}\, (2\pi)^3\delta^3(p-k)
\nonumber\\&&\times 
h(p_1^2)h(p_2^2) h(k_1^2)h(m^2) \, ,\qquad
\label{eq:VC}
\eea
where $p_1=p+P/2$, $p_2=p-P/2$ (and similarly for $k_1$ and $k_2$),
$C$ is a constant, $h$ is the strong form factor that models the quark-gluon vertex, and in this example  $\hat k_2^2=m^2$, with $m$ the dressed quark mass.  The chiral limit of  (\ref {eq:VC}) follows by setting $m\to m_\chi$, $P\to0$, and $h(m_\chi^2)=1$.  

Returning to Eq.~(\ref{eq:GinBSform}), extracting the $\gamma^5$,  and using the chiral limit of (\ref{eq:VC}) gives
\bea
G(p^2)&=&\frac{C}{m_\chi}\, h^2(p^2)\int d^3k\, \delta^3(p-k)G(\hat k^2) 
\nonumber\\
&=&\frac{C}{m_\chi}\,  h^2(p^2) G_0 \label{eq:Gchiral}
\eea
where $\hat k=\{E_k,{\bf k}\}$ is the value of the four-vector $k$ at the spectator pole, and the second line 
employs the definition of the chiral limit of the vertex function, $G(\hat k^2)\equiv G_0$.  Note that placing the external particle on-shell gives a consistent equation only if $C=m_\chi$, which is another way of showing the constraint on $C$ in the chiral limit that was discussed in Ref.~I.

The result (\ref{eq:Gchiral}) suggests that, near the chiral limit,  the vertex functions with one particle on-shell should be well approximated by
\begin{eqnarray}
 \Gamma(p_1,\hat p_2)&=&\gamma^5 h(p_1^2) G_0
 \nonumber\\
 \Gamma(\hat p_1, p_2)&=&\gamma^5 h(p_2^2) G_0 \,.
 \label{eq:redvertex}
\end{eqnarray}
The validity of this approximation depends on the observation that the most rapid variation of the scalar functions that define $\Gamma(p_1,p_2)$ is through its dependence on the strong form factors $h$.  

\subsection{Off-shell quark current}

In order to calculate a conserved current for processes involving bound states, we employ the general framework introduced by Riska and Gross \cite{Gro87}.  Here the strong form factors ($h$ in this paper) attached to the interaction vertices are moved to the propagators connecting neighboring vertices, where they provide an additional modification of the dressed quark propagators connecting two bare vertices.  Consistency then requires that these form factors also be \lq \lq factored out''  of the quark current, leading to the introduction of a 
\emph{reduced} or \emph{bare} electromagnetic current for the off-shell quarks defined by
\begin{eqnarray}
j^\mu_R (p',p)=h^{-1}(p'^2)j^{\mu} (p',p)h^{-1}(p^2)\,.
\end{eqnarray}
In order to ensure current conservation~\cite{Gro87,Gro96}, $j^\mu_R$ must satisfy the Ward-Takahashi (WT) identity:
\begin{eqnarray}
 q^\mu j_{R\mu } (p',p)=\widetilde S^{-1}(p)-\widetilde S^{-1}(p')\,,\label{eq:WTI}
\end{eqnarray}
where $\widetilde S(k)$ is the dressed quark propagator multiplied by the square of  the quark form factor 
\begin{eqnarray}
 \tilde S(p)=h^2(p^2) S(p)\,.
\end{eqnarray}
The simplest form of the reduced  current that can satisfy the WT identity (\ref{eq:WTI}) with a dressed propagator with a {\it mass function depending on momentum\/} is a generalization of the current previously introduced in Ref.~\cite{Gro96}
 \begin{eqnarray}
&& j_{R}^\mu (p',p)
\nonumber\\&&
\quad=f(p',p)
% \nonumber\\&&\times
 \bigg[{\cal G}_1^\mu(q)
 %\nonumber\\&&
  +\kappa F_2(q^2)\frac{\mathrm i \sigma^{\mu\nu}q_{\nu}}{2m}\bigg]
 \nonumber\\&&\qquad+ \delta(p',p)\Lambda(-p')\,{\cal G}_4^\mu(q)+\delta(p,p'){\cal G}_4^\mu(q)\,\Lambda(-p)
  \nonumber\\&&\qquad+ g(p',p)\Lambda(-p')\,{\cal G}_3^\mu(q)\,\Lambda(-p)\,,
\end{eqnarray}
where $\Lambda(-p)=(M(p^2)-\slashed{p})/2M(p^2)$, and (for $i=1,3,4$)
\bea
{\cal G}^\mu_i(q)\equiv  \Big(F_i(q^2)-1\Big)\widetilde\gamma^\mu+\gamma^\mu\, .
\eea
Here the transverse gamma matrix, $\widetilde \gamma^\mu=\gamma^\mu-q^\mu \slashed{q}/q^2$, makes no contribution to the WT identity,  and the $F_i(q^2)$ (with $i=1, \ldots,4$) are dressed quark form factors (including two new off-shell form factors $F_3$ and $F_4$).  All of the quark form factors are constrained by $F_i(0)=1$ with $\kappa$  the anomalous magnetic moment of the quark. 
The functions $f$, $g$, and $\delta$ are fixed by the requirement that $j_{R}^\mu$ satisfies the WT identity~(\ref{eq:WTI}).   Using the notation $h=h(p^2)$, $h'=h(p'^2)$, $M=M(p^2)$, and $M'=M(p'^2)$, with a propagator
\bea
\tilde S^{-1}(p)=\frac{M-\slashed{p}}{h^2}\, .
\eea
a short calculation yields
\begin{eqnarray}
  g(p',p)&=&\frac{4MM'}{h^2h'^2}\frac{(h^2-h'^2)}{(p'^2-p^2)} \\
  \delta(p',p)&=&\frac{2M'}{h'^2}\frac{(M'-M)}{(p'^2-p^2)} \\
  f(p',p)&=&\frac{M^2-p^2}{h^2(p'^2-p^2)}-\frac{M'^2-p'^2}{h'^2(p'^2-p^2)}\,.
  \end{eqnarray}
Note that, if $M'=M$, $\delta$ vanishes and $f$ and $g$ reduce to results previously given in the literature.  When contracted into a conserved current, or a physical photon, the terms proportional to $q^\mu$ vanish, reducing ${\cal G}_i^\mu(q)$ to
\bea
&&{\cal G}_i^\mu(q)\to F_i(q^2)\gamma^\mu\,.
\eea

The four quark form factors $F_i$ can be calculated in the CST, but  this exercise will be saved for another day.    For now we will use the quark current in the chiral limit, where the mass function reduces to \cite{mass_function_paper}
\bea
M_\chi(p^2)=m_\chi h^2(p^2) \, ,\label{eq:chiralmf}
\eea
and, as appropriate for a point-like bare quark, $\kappa=0$ and all form factors are set to unity.  This simplifies the off-shell structure functions
\bea
g_\chi(p',p)&=&-2\delta_\chi(p',p)=\frac{4m_\chi^2(h^2-h'^2)}{(p'^2-p^2)}\\
f_\chi(p',p)&=&\frac14g_\chi(p',p)+\frac{p'^2h^2-p^2h'^2}{h^2h'^2(p'^2-p^2)}
\nonumber\\
&=&\frac{M_\chi^2-p^2}{h^2(p'^2-p^2)}-\frac{M'^2_\chi-p'^2}{h'^2(p'^2-p^2)}
\eea
and reduces the current to
\bea
j^\mu_{R_\chi}(p',p)&=&\frac{[\slashed{p}'\gamma^\mu\slashed{p}+p'^2\gamma^\mu]}{h'^2(p'^2-p^2)} -\frac{[\slashed{p}'\gamma^\mu\slashed{p} +p^2\gamma^\mu]}{h^2(p'^2-p^2)}
\nonumber\\
&=&\gamma^\mu\frac{(h^2p'^2-h'^2p^2)}{h^2h'^2(p'^2-p^2)} +\frac{\slashed{p}'\gamma^\mu\slashed{p}}{h^2h'^2}\frac{(h^2-h'^2)}{(p'^2-p^2)} \, .\qquad
\label{eq:reducedcurr}
\eea

It is interesting to compare this with the Ball-Chiu (BC) \cite{BC80} current used by Maris and Tandy~\cite{MT00}.  In our notation, denoting $p'+p=2P$  their current is
\bea
j^\mu_{\rm BC}(p',p)&=&\gamma^\mu \frac{h^2+h'^2}{2h^2h'^2}
%\nonumber\\&&
+\frac{2\slashed{P}P^\mu}{h^2h'^2}\frac{h^2-h'^2}{p'^2-p^2}\, .
\eea
The difference between these two currents is 
\bea
j^\mu_{R_\chi}(p',p)-j^\mu_{\rm BC}(p',p)=\frac{X^\mu }{2h'^2h^2}\frac{(h'^2-h^2)}{(p'^2-p^2)}
\eea
where $X^\mu$ is purely transverse  
\bea
X^\mu=\slashed{p}'\gamma^\mu\slashed{p}=\slashed{p}\gamma^\mu\slashed{p}'-\slashed{q}q^\mu+q^2\gamma^\mu\, .\qquad
\eea
The fact that $X^\mu\ne0$ is a demonstration that the current cannot be uniquely determined by the WT identity alone.  It was only after we derived our current that we became aware of the BC current used by Maris and Tandy. In fact, Maris and Tandy used this freedom to add a transverse  $\rho$ contribution to their current. In the absence of a dynamical calculation, we know of no way to determine these transverse contributions. 

In any case, the quark current can be computed from an integral equation that sums the $q\bar q$ interaction to all orders, and includes automatically contributions from the $\rho, \rho', \cdots$ tower of vector meson states.   Since our formalism can be applied equally well to the time-like region where these states live, this will be explored in the near future.

 \section{Final steps}

\subsection{Reduction of the current}

Substituting (\ref{eq:redvertex}) and (\ref{eq:reducedcurr}) into the pion current (\ref{eq:picurrentA}) gives

  \begin{eqnarray} 
\label{eq:pioncurrent3}
 && F_\pi(Q^2)2P_0^0
\nonumber\\&&
\quad=- G_0^2\,\int_k 
 \mathrm {tr} \Big[\widetilde S(p_+)  j_{R_\chi}^0 (p_+,p_-)\widetilde S(p_-) 
% \nonumber\\&&\qquad\qquad\qquad\qquad\times
\Lambda (\hat k)\Big],  \qquad
\end{eqnarray}
where we keep the pion mass $\mu\ne0$, but take the chiral limit elsewhere so that the on-shell quark has mass $m_\chi$ (so that $E_k$ is now $\sqrt{m_\chi^2+{\bf k}^2}$), the $\gamma^5$s have been removed,  
and we specialize to the Breit frame~(\ref{eq:BreitFrame}) where
\bea
p^2_\pm&=&\mu^2+m_\chi^2-2P_0E_k\pm k_zQ\,.
\eea 
 Evaluating the trace gives
\bea
F_\pi(Q^2)=-\frac{G^2_0}{P_0}\int_k \frac{{\cal N}}{\mathcal D} \, ,
\label{eq:FFint}
\eea
where, using the notation $M_\pm=M_\chi(p_\pm^2)$ and $h_\pm=h_\chi(p_\pm^2)$,  
\bea
&&{\cal N}=-2h_\Delta\Bigg\{4P_0^2E_k^3-2P_0E_k^2(M_+M_-+3 \bar m^2 -2M_S)
\nonumber\\
&&+E_k\Big[2 \bar m^2 (M_+M_- + \bar m^2)-2m_\chi M_S(2P_0^2+\bar m^2)
\nonumber\\&&
\qquad+4m_\chi^2P_0^2-k_z^2Q^2+m_\chi k_z Q  M_\Delta\Big]
\nonumber\\&&
-m_\chi P_0\Big[2m_\chi(M_+M_-+\bar m^2)-2\bar m^2 M_S +k_z Q M_\Delta \Big]\Bigg\}
\nonumber\\
&&+2 k_z Q \, h_S \Big[m_\chi P_0(M_S-2m_\chi)
\nonumber\\&&
\qquad\qquad+E_k(M_+M_- +\bar m^2-m_\chi M_S)\Big]
\eea
with $h_S=h_++h_-$, $h_\Delta=h_+-h_-$,  $M_S=M_++M_-$,  $M_{\Delta}=M_+-M_-$, and $\bar m^2=m_\chi^2+\mu^2$.  The denominator is
\bea
\mathcal D&=&2k_zQ(M_+^2-p_+^2)(M_-^2-p_-^2) \, .
\eea

To study the convergence of these integrals, look at the limit as $k$ becomes very large (in this section we use the notation $k \equiv |{\bf k}|$). In this limit, the running quark masses can be neglected compared to factors of $k$, giving
\bea
{\cal N}&\stackrel{k\gg m_\chi}{\longrightarrow}&2k(h_+^2-h_-^2)\Big[k_z^2 Q^2-4k^2P_0^2\Big]
\nonumber\\
\mathcal D&\stackrel{k\gg m_\chi}{\longrightarrow}& 2k_zQ(4P_0^2k^2-k_z^2Q^2)\, .
\eea 
Ignoring details, the most divergent term therefore behaves like
\bea
\int_k\frac{{\cal N}}{\mathcal D}&\simeq&\int_k h^2 \, ,
\\
\nonumber\\ &&\nonumber
\eea
%
%\vspace{0.1in}
guaranteeing that the integrals will converge if
\bea
\lim_{k\to\infty}h^2<\frac1{k^2}\, . \label{eq:conv}
\eea
This requires that $h$ approach zero faster than $k^{-1}$.

\subsection{High-$Q^2$ limit}
\label{sec:highQ2limit}

The high-$Q^2$ limit of the form factor is particularly interesting.  In preparation for this discussion, use the symmetry of the integrand to convert (\ref{eq:FFint}) to
\bea
\int_{-\infty}^\infty \, dk_z {\cal F}(k_z)=2\int_{0}^\infty dk_z \,{\cal F}(k_z)\, .
\eea 
Next, study the arguments $p_+^2$ and $p_-^2$ in the limit of large $Q$.  Keeping terms to order $1/Q$ in $p_+^2$ (which will turn out to be sufficient), gives
\bea
p_+^2&\to& \mu^2+m_\chi^2-Q(E_k-k_z)-\frac{2\mu^2}{Q}E_k
\nonumber\\
p_-^2&\to& -Q(E_k+k_z)\, . \label{eq:ppmQ}
\eea
Note that, as $Q \rightarrow \infty$, $p_-^2\to -\infty$ for all values of $k_z$, and hence only the leading term is needed.  The functions $h_-$ and $M_-$ vanish in that limit. In contrast, the behavior of $p_+^2$ at large $Q$ depends on the size of $k_z$.  The approximate formula (\ref{eq:ppmQ}) shows that $p_+^2\to -\infty$ at the limits of the $k_z$ integration.  We saw already in the discussion of Eq.~(\ref{eq:limI})  that the integrand only deviates significantly  from zero in the vicinity of  the critical value
\bea
k_{z0}\simeq \frac{Q E_\perp}{2\mu} 
\eea
for which  $p_+^2$ reaches its maximum, and where $E_\perp=\sqrt{m_\chi^2+k_\perp^2}$.  For large, finite $Q$, this is a large value of $k_z$ that cannot be ignored.  At this critical point, 
\bea
E_k&\to&\frac{QE_\perp}{2\mu}\Big(1+\frac{2\mu^2}{Q^2}\Big) \, ,
\nonumber\\
p_+^2&\to& p_c^2=\mu^2+m_\chi^2- 2 \mu E_\perp\, .
\eea

\begin{figure}[t]
\begin{center}
\includegraphics[width=0.45\textwidth]{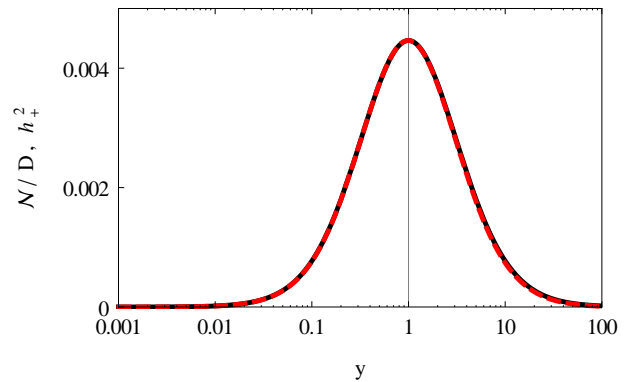}
\caption{(Color online) ${\cal N}/\mathcal D$ times a constant (black solid line) compared with $h_+^2$ (red dashed line) for large $Q^2$. Both curves lie on top of each other and they are strongly peaked at $y=1$.}\label{fig:integrand_peak}
\end{center} 
\end{figure}

To understand the integral it is convenient to introduce the momentum fraction
\bea
x\equiv \frac{E_k-k_z}{2E_k}
\eea
The $k_z$ integration will be replaced by an integration over $x$ and since $k_z>0$ the $x$-integration varies between 0 and $\frac12$, with the Jacobian
\bea
\int_0^\infty \frac{dk_z}{E_k}=\int_0^\frac12  \frac{dx}{2x(1-x)}\, .
\eea
In terms of this variable,
\bea
E_k&=&\frac{E_\perp}{2\sqrt{x(1-x)}}
\nonumber\\
k_z&=&\frac{E_\perp(1-2x)}{2\sqrt{x(1-x)}}
\nonumber\\
p_+^2&\to&\mu^2+m_\chi^2-\frac{Q E_\perp}{\sqrt{x(1-x)}}\Big(x+\frac{\mu^2}{Q^2}\Big)\, .
\eea
The last expression displays clearly how the non-leading term is needed to give the correct limit $p_+^2\to-\infty$ as $x\to0$, and that $p_+^2$ is finite (and hence $h_+^2$ large) only in the limited region of small $x\sim \mu^2/Q^2$.  Hence we may assume $x\ll1$ and write
\bea
p_+^2\to \mu^2+m_\chi^2 -\mu E_\perp\Big(\sqrt{y}+\frac1{\sqrt{y}}\Big)
\eea
with $y=x (Q/\mu)^2$.  In terms of the variable $y$, the integrand peaks near $y\sim1$, as shown in Fig.~\ref{fig:integrand_peak}.    
Evaluation of the form factor at high $Q$ is therefore ideally suited to a peaking approximation, with the slowly varying part of the integral evaluated at the peak.
Fig.~\ref{fig:integrand_peak} compares  $h_+^2$ with the integrand, $\mathcal N/\mathcal D$, scaled by a constant factor.  Both curves lie on top of each other showing that the peaking approximation works very well.
With these approximations, the form factor at large $Q$ becomes 
\bea
F_\pi(Q^2)\stackrel{Q^2\gg\mu^2}{\simeq}-\frac{2G_0^2}{Q}\int_{k_\perp}\int_0^\infty \frac{dy}{y} h^2(p_+^2) \Big[\frac{{\cal \tilde N}}{\mathcal D}\Big]_{\rm peak}\, .\qquad
\eea
Evaluation of the terms at the peak gives
\bea
{\cal \tilde N}&\to&-Q^3\frac{2 E_\perp}{\mu^2}\Big( E_\perp^2 \mu-2 E_\perp (m_\chi^2-m_\chi M_++\mu^2)\nonumber\\&&+2 m_\chi \mu (m_\chi-M_+)\Big)
\nonumber\\
\mathcal D&\to& Q^4 \frac{E_\perp^2}{\mu^2}\Big(M_+^2-\mu^2-m_\chi^2+2\mu E_\perp\Big)\, .
\eea
Hence the form factor falls like $Q^{-2}$ at large $Q$, with the coefficient independent of the detailed structure of the strong form factor $h$.  

\section{Results} \label{sec:Resuts}

\begin{figure}
    \includegraphics[height=5.5cm]{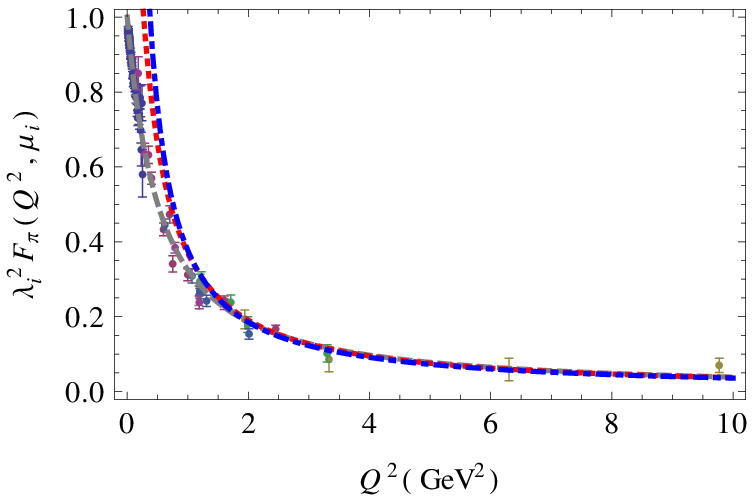} \\\vspace{0.5cm}
     \includegraphics[height=5.5cm]{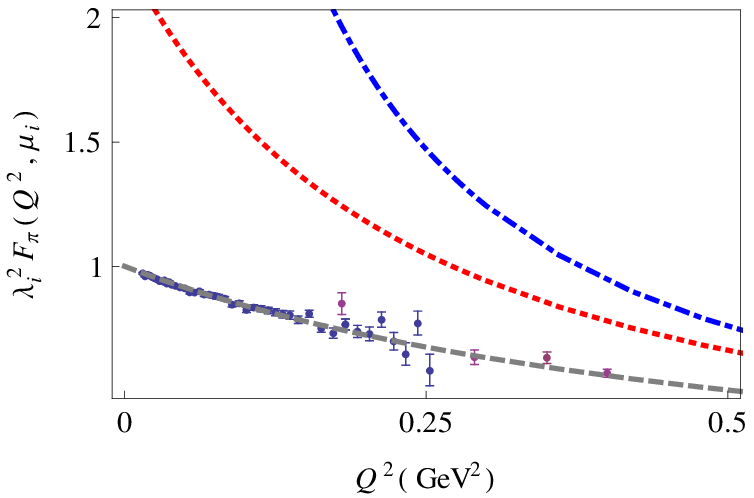} 
        \caption{(Color online)
                        The pion form factor $F_\pi (Q^2,\mu_i)$ scaled with $\lambda_i^2= (\mu_1/\mu_i)^2$ for different pion masses $\mu_1=0.42$ GeV (gray dashed line), $\mu_2=0.28$ GeV (red dotted line) and $\mu_3=0.14$ GeV (blue dotdashed line) compared with the data~\cite{Amendolia:1986wj,Brown:1973wr,Bebek:1974iz,Bebek:1976ww,Bebek:1978pe,Volmer:2000ek,Horn:2006tm,Tadevosyan:2007yd,Huber:2008id}, at high $Q^2$ (top) and low $Q^2$ (bottom).} \label{fig:piff}
\end{figure}

The numerical results  for the pion form factor presented in this paper use the simple strong quark form factor $h(p^2)$,
\begin{eqnarray}
 h(p^2)=\left(\frac{\Lambda_\chi^2-m_\chi^2}{\Lambda_\chi^2-p^2}\right)^n\,, \label{eq:hff}
\end{eqnarray}
obtained in Ref.~I.  Here $\Lambda_\chi=2.04$ GeV is a mass parameter  determined by a fit of the quark mass function to the lattice QCD data.  The power $n=2$  is not inconsistent with the lattice data and ensures that the integrals will converge. 
Note that $h(p^2)$ has a pole at $\Lambda_\chi^2=p^2$, but this point lies far outside of the region of the $k$ integration.

Our pion form factor is very insensitive to the particular choice of $h$, as long as $n>1$ for convergence. This remarkable property can be understood, at least for large $Q^2$, from the analysis of Sec.~\ref{sec:highQ2limit}, which revealed that the high-$Q^2$ behavior of the form factor integral is completely determined by its integrand evaluated at the peaking value $k_z=k_{z0}$ of $h$.  In this work we have neglected the anomalous moment term in the quark current,  proportional to $\kappa$.  Conventional arguments  suggest that it should be small at large $Q$ and it would vanish for point-like quarks.

We emphasize that our model, in its present form, gives reliable results only for pion masses in a limited range.  
In particular, if $\mu$ is  larger than the threshold value of $\mu_s=2m_\chi$, the dressed quark propagators develop poles, which allow \emph{both} quarks to go on-mass-shell \emph{at the same time} [recall the discussion following Eq.~(\ref{eq:limI})].  This can happen only because we have not yet included the confining part of the interaction.  Once confinement is included this cut will vanish.  
Since the value $\mu_s=2m_\chi$ is far above the physical pion mass this does not represent a serious limitation of the present model.

For values of $\mu$ below the  threshold value $2m_\chi$, we showed in Sec.~\ref{sec:FFandCST} that the RIA  used in this paper breaks down for small pion masses at small $Q^2$.  This happens because the pole contributions from the struck quark, neglected in RIA, become large.  
Therefore, the form factor in RIA for a physical pion mass is reasonable at large $Q^2$, but  too large at small $Q^2$ and does not give the correct charge at $Q^2=0$.   
Therefore, for values of $\mu$ somewhere near the chiral quark mass, $m_\chi$, the RIA is a good approximation.  

Since the pion form factor depends on the pion mass, we adopt the notation $F_\pi(Q^2,\mu)$.  In all cases, $F_\pi(0,\mu)=1$. 
We found that the value $\mu$=0.42 GeV gave the best fit to the data over the full range of $Q^2$,  so we adopted this form factor as a standard of comparison (whenever we do not explicitly specify the pion mass in the form factor argument, the value $\mu$=0.42 GeV is implied).

We find a remarkable scaling behavior at large $Q^2$:
\begin{eqnarray}
   F_\pi(Q^2,\lambda \mu)\stackrel{Q^2\gg\mu^2}{\simeq}\lambda^{2} F_\pi(Q^2,\mu)\,, \label{eq:scaling}
\end{eqnarray} 
 where $\lambda$ is a scaling parameter.   In particular, Fig.~\ref{fig:piff} shows the form factor results for $\mu_3=0.14$~GeV and $\mu_2=0.28$~GeV when scaled to fit the result for $\mu_1=0.42$~GeV. The three curves are compared with the experimental data~\cite{Amendolia:1986wj,Brown:1973wr,Bebek:1974iz,Bebek:1976ww,Bebek:1978pe,Volmer:2000ek,Horn:2006tm,Tadevosyan:2007yd,Huber:2008id}.

\begin{figure}
  \includegraphics[height=5.5cm]{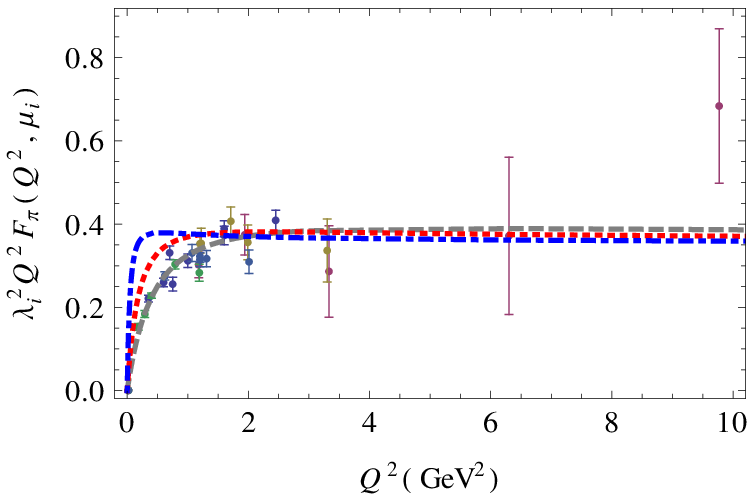} 
  \\\vspace{0.5cm}
   \includegraphics[height=5.5cm]{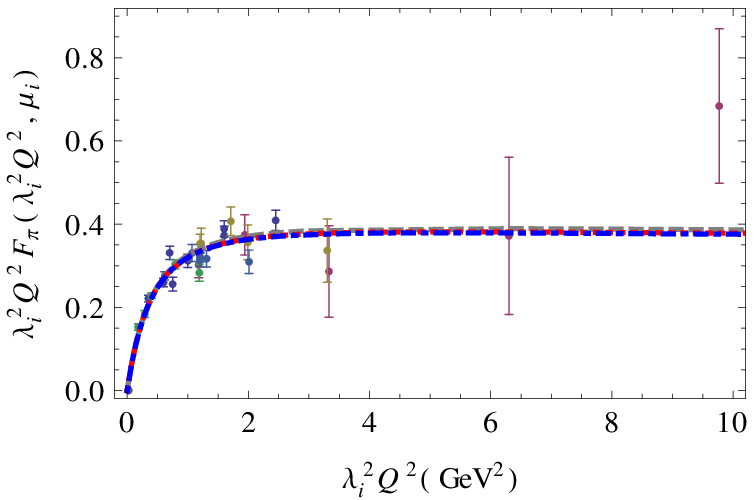} 
        \caption{(Color online) Scaled form factors compared to the data~\cite{Amendolia:1986wj,Brown:1973wr,Bebek:1974iz,Bebek:1976ww,Bebek:1978pe,Volmer:2000ek,Horn:2006tm,Tadevosyan:2007yd,Huber:2008id}.  The top panel shows $\lambda_i^2 Q^2 F_\pi (Q^2,\mu_i )$; the bottom shows $\lambda_i^2 Q^2 F_\pi (\lambda_i^2 Q^2,\mu_i)$.  In both panels $\lambda_i=\mu_1/ \mu_i$for different pion masses $\mu_1=0.42$ GeV (gray dashed line), $\mu_2=0.28$ GeV (red dotted line) and $\mu_3=0.14$ GeV (blue dotdashed line). } 
\label{fig:Q2piff}
\end{figure}

As $Q^2$ becomes smaller the three curves diverge as a consequence of the breakdown of the RIA in the small-$Q^2$ region for small $\mu$.  In the high-$Q^2$ region, as shown in the top panel of Fig.~\ref{fig:Q2piff},  the curves almost lie on top of each other.   Furthermore, if the $Q^2$ dependence is also scaled over the whole range of $Q^2$
\begin{eqnarray}
F_\pi(\lambda^2 Q^2,\lambda \mu)\simeq F_\pi(Q^2,\mu) \, \label{eq:scaling2}
\end{eqnarray}
the curves almost lie on top of each other, even at small $Q^2$.  This expresses the fact that our form factor depends very weakly on the remaining scale-dependent quantities, $m_\chi$ and $\Lambda_\chi$.

The form factors satisfy a nearly monopole behavior
 \begin{eqnarray}
F_\pi(Q^2)\stackrel{Q^2\gg\mu^2}{\sim} \frac{ 1}{Q^2+\nu^2}\, \label{eq:monoploe}
\end{eqnarray}
with a mass scale of $\nu\simeq 0.63$ GeV obtained from a fit to the $\mu_1=0.42$ GeV form factor. 
\begin{figure}
  \includegraphics[height=5.5cm]{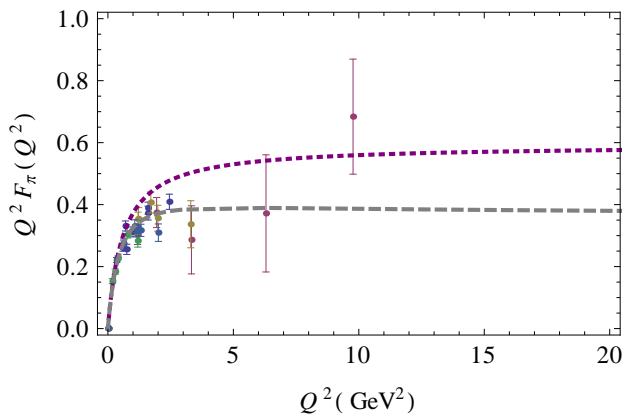} 
 \caption{(Color online) A $\rho$-pole contribution to the pion form factor, normalized to unity at $Q^2=0$ (purple dotted line), compared to our calculation with $\mu=0.42$ GeV (gray dashed line).} \label{fig:Rho}
\end{figure}

\section{Summary and conclusions}\label{sec:conclusions}

This paper uses the Covariant Spectator Theory (CST) to compute the pion form factor in the relativistic impulse approximation (RIA).  The CST is formulated  in Minkowski space, so that even though  results for the form factor at space-like momentum transfer ($q^2=-Q^2<0$) are presented here, the theory can be used to calculate the form factor in the time-like region ($q^2>0$)  as well.   The manifestly covariant dynamical model for the $q\bar q$ interaction that is the foundation of the calculations presented here incorporates both spontaneous chiral symmetry breaking and confinement, and it is discussed in Ref.~I.  Some features of this model were previously introduced by Gross, Milana and \c{S}avkli \cite{QQbar, Savkli:1999me}.

This first calculation of the pion form factor  uses the quark mass function obtained in Ref.~I,   and expresses the pion vertex function in terms of this mass function.  This approximation is particularly good near the chiral limit.   
We emphasize that this is a very simple picture for the pion. Still, when combined with the results of Ref.~I, we show that this simple picture can give results that are in good agreement with both the lattice data for the dressed quark mass and the experimental data for the pion electromagnetic form factor.   We find some interesting scaling relations relating form factors  with different values of $\mu$.

An interesting issue remains.  This simple model is able to describe the data well, yet seems to include no contribution from the $\rho$ meson that is expected from vector meson dominance.  (For a comparison of our model with what is expected from a simple $\rho$ pole, see Fig.~\ref{fig:Rho}.)  Where is the $\rho$ contribution?  It should be contained in the dressing of the quark current, $j^\mu$.   Maris and Tandy~\cite{MT00} suggest that their Ball-Chiu current contains some of these contributions (in which case our dressed current probably also contains them).  The balance between the triangle diagram with no $\rho$ contribution and contributions coming from the dynamical dressing of the quark current, including the $\rho$ pole, will best be understood once the dressed quark current has been calculated in both the time-like and space-like regions.  

For a more quantitative study of the light-meson properties the solution of the complete four-channel CST equation and a fit to the light-meson spectrum is needed, which will be the subject of our future program.

\begin{acknowledgements}
T.P. is pleased to acknowledge valuable discussions with Gernot Eichmann. This work received financial support from Funda\c c\~ao para a Ci\^encia e a Tecnologia (FCT) under Grants No.~PTDC/FIS/113940/2009, No. CFTP-FCT (PEst-OE/FIS/U/0777/2013) and No. POCTI/ISFL/2/275. This work was also partially supported by the European Union under the HadronPhysics3 Grant No. 283286, and by Jefferson Science Associates, LLC under U.S. DOE Contract No. DE-AC05-06OR23177.
\end{acknowledgements}

\bibliographystyle{h-physrev3}
\bibliography{PapersDB-v1.4E}
\end{document}